\documentclass[runningheads]{llncs}
\usepackage{graphicx}
\graphicspath{{./}}
\usepackage{caption}
\usepackage{subcaption}
\begin{document}
\title{UniquID: A Quest to Reconcile Identity Access\\ Management and the Internet of Things}
%
%
\author{Alberto Giaretta\inst{1} \and
Stefano Pepe\inst{2} \and
Nicola Dragoni\inst{1,3}}
\authorrunning{A. Giaretta et al.}
%
\institute{Centre for Applied Autonomous Sensor Systems (AASS), \"Orebro University, \"Orebro, Sweden.
\email{alberto.giaretta@oru.se} \and
UniquID Inc., San Francisco, USA.
\email{pepe@uniquid.com} \and
DTU Compute, Technical University of Denmark, Kongens Lyngby, Denmark
\email{ndra@dtu.dk}}
\maketitle              
\begin{abstract}
The Internet of Things (IoT) has caused a revolutionary paradigm shift in computer networking. After decades of human-centered routines, where devices were merely tools that enabled human beings to authenticate themselves and perform activities, we are now dealing with a device-centered paradigm: the devices themselves are actors, not just tools for people. 
Conventional identity access management (IAM) frameworks were not designed to handle the challenges of IoT. Trying to use traditional IAM systems to reconcile heterogeneous devices and complex federations of online services (e.g., IoT sensors and cloud computing solutions) adds a cumbersome architectural layer that can become hard to maintain and act as a single point of failure.
In this paper, we propose UniquID, a blockchain-based solution that overcomes the need for centralized IAM architectures while providing scalability and robustness. 
We also present the experimental results of a proof-of-concept UniquID enrolment network, and we discuss two different use-cases that show the considerable value of a blockchain-based IAM.
\end{abstract}
\section{Introduction}\label{sec:trad_iam}
Information Technology (IT) has radically changed throughout history. In just a few decades, computers evolved from bulky, standalone machines that filled rooms to small and powerful devices capable of gathering information from other devices over the Internet. This explosion of capabilities has led to a concomitant expansion of complexity in a variety of areas, including identity access management (IAM).

Conventional IAM systems are essential for traditional local networks and businesses, but not well-suited for large networks of complex, highly distributed devices, such as the combination of Internet of Things (IoT) and machine to machine (M2M) communication. The main reason is that IAM systems were designed for human beings, not devices. Until a decade ago, a typical scenario involved a known number of terminals and a comparable number of human users with well-defined roles. Accounts were issued for each person and access rights stored on a central server, making access control (AC) relatively easy to manage. Today, IT is much more complex than it once was. 

The Internet of Things is causing the old user-centered paradigm to shift toward a device-centered one. Previously, accounts were tied to human beings and devices were just a means to the end of accomplishing all of the tasks those humans performed. Today, devices are increasingly becoming the actors themselves, with the tasks narrowed down from general purpose functions to very specific operations. 
With this in mind, traditional IAM solutions are expensive, make maintaining cross-domain consistency challenging, and represent a critical single point of failure in organizations due to their centralized nature.

\begin{figure}[ht]
\center
\includegraphics[trim=0mm 0mm 0mm 0mm,clip,width=0.75\textwidth]{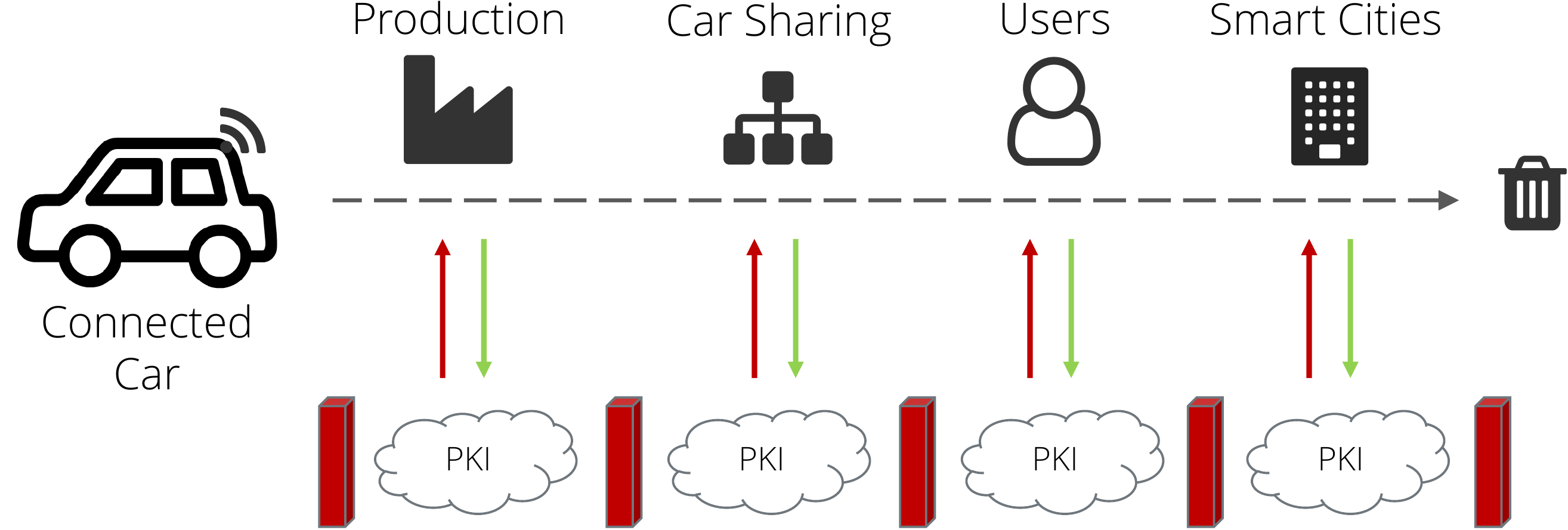}
\caption{PKIs inherently introduce compartmentalization, which increases the complexity of identity reconciliation over cross-domain federated architectures. In this example, the identity of a single Internet-enabled car has to be separately managed by several different PKIs.}
\label{fig:pki_compartments}
\end{figure}

In the 1990s, PKI was considered one of the key technologies for dealing with security issues and enabling trust among parties~\cite{carayannis2006innovation}. But adoption of the technology never took off as envisioned, due to critical issues such as privacy and liability concerns, management complexity, and high costs. 
PKI technology is compartmentalized by design, as shown in Figure~\ref{fig:pki_compartments}, and therefore irremediably ill-suited for devices that need cross-domain interactions or technical solutions that rely upon the federation of independent services. PKI lacks the flexibility needed for IoT and M2M, whereas certificate-based authentication is well-suited for these uses. 

This paper describes UniquID~\footnote{http://www.uniquid.com}, a solution based on an infrastructure that takes advantage of certificates and blockchain technology to overcome the difficulties in reconciling IoT credentials and cross-domain IAM. The goals of this paper include the following:
\begin{itemize}
\item Describing a cheaper and simpler alternative to traditional IAM systems.
\item Illustrating the implementation of cross-domain identities for IoT devices to circumvent account reconciliation.
\item Showing how the proposed design removes single points of failure from the trust structure.
\item Demonstrating direct peer-to-peer (P2P) authentication and authorization among  IoT devices.
\item Showing how an IoT device, empowered to locally confirm a smart transaction, deals with partitioning issues as defined by the CAP Theorem. 
\end{itemize}

\subsection{Paper Outline}
The paper is organized as follows. Section~\ref{sec:rel_work} briefly presents the main players in the blockchain-based IAM landscape, how UniquID addresses some of the distributed IoT challenges, and how it is linked to the CAP Theorem. Section~\ref{sec:overall_architecture} describes the general concepts behind UniquID, and 
Section~\ref{sec:evaluation} shows the experimental evaluation of an enrolment proof-of-concept. Section~\ref{sec:case_studies} describes two example case studies, and Section~\ref{sec:concl} lays out the paper conclusions and discusses future work.

\section{Related Work}\label{sec:rel_work}
In the past, blockchain has already been advocated for more secure IAM systems. For example, Kshetri~\cite{kshetri_2017} suggests that blockchain can help in strengthen the IoT in different ways, such as preventing DDoS and IP spoofing attacks. Furthermore, Gartner estimates that by 2020 the IoT will require up to 1000 times the 2016 network capacity~\cite{gartner_2015}. This entails that the centralized IAM paradigm might not scale enough to tolerate such requirements, and this is where the decentralized IAM can provide both more security and scalability.

In Roman et al.~\cite{ROMAN20132266}, the authors lay out the main challenges of distributed IoT. They argue that identity and authentication are primary concerns due to the inherent dynamism introduced by device mobility, unstable connections, and related problems. Throughout this paper, we show how UniquID provides direct identification and authentication, which in turn enables efficient M2M resource negotiation. Another issue raised is security, which UniquID ensures through asymmetric encryption, adding symmetric cipher-based encryption for larger data streams. Furthermore, depending on the chosen backbone 
, privacy can be partially or even totally lost. In cases where privacy is needed, it can be enforced through non-interactive zero-knowledge proofs such as zk-SNARKs~\cite{zk_snarks} and Bulletproofs~\cite{bulletproofs}, but this involves considerable overhead that may not be appropriate for resource-constrained devices.

Several parallel projects have been proposed to address the problem of identity management over blockchain, yet most of them do not address the access management part. For example, IBM Hyperledger Indy aims to provide an SDK solution to manage identities over distributed ledgers~\cite{dhillon2017hyperledger}. Even though UniquID similarly provides an SDK, our proposal addresses both the identity and access management parts of the equation, whereas Indy addresses only identity.

Scholars proposed some solutions as well. Le and Mutka propose CapChain~\cite{le_2018}, a blockchain-based access control framework that enables IAM on public blockchains, ensuring at the same time privacy; in particular, the authors built a proof-of-concept over Monero source code and ran processing time local benchmarks. Ouaddah et al.~\cite{ouaddah_2017} propose a blockchain-based access control named FairAccess, and implemented a proof-of-concept which uses Bitcoin OP\_RETURN as a storage field. Finally, Novo~\cite{novo_2018} proposes a blockchain-based access management that utilizes a management hub (e.g., an edge node) as a middlepoint between IoT devices and the blockchain.

Even though all the aforementioned papers prove that blockchain-based IAM is possible, they lack a thorough discussion on how the underlying blockchain choice can deeply affect operational costs, data immutability, immediacy, and efficiency. Moreover, none of such works envisage a way to verify a policy without Internet connection, failing to take full advantage of the M2M paradigm. In particular, no one highlights the strong relationship between Brewer's CAP Theorem and IAM solutions, nor motivations and implications of choosing consistency over availability (or vice-versa) in their proposals.

\subsection{Consistency, Availability, and Partition Tolerance (CAP)}
According to Brewer's CAP Theorem~\cite{Brewer2012}, one cannot ensure all three of the following in a system at any given time: \textit{consistency (C)}, \textit{availability (A)} and \textit{partition tolerance (P)}. CAP is often misunderstood, with people thinking that a distributed system is always unable to assure all three requirements. In reality, the choice is only between consistency and availability when a partition or a failure occurs; under normal circumstances, all three can be assured simultaneously. Another misunderstanding is about consistent-available (CA) systems, which are simply not possible in a distributed scenario. According to CAP, a system could be designed to be CA, but it would require a network that ensures no packet is ever dropped at any moment in time. For a fixed partition tolerance requirement, the only real choice is between consistency and availability.

Consistency, which is a property related to read operation, can be either strong or eventual. In a consistent-partition-tolerant (CP) scenario typical of an RDBMS, the system ensures that every commit to the database is propagated and kept consistent throughout all database replicas, so that every read operation returns the most recently updated result. In an available-partition-tolerant (AP) scenario typical of NoSQL, the read operation does not ensure that the user receives the most up-to-date result. But even though AP sounds problematic, consistency is eventually achieved, and this approach is common in many non-critical applications due to its strong support for partition tolerance and availability. 

It is important to keep in mind that partitions and latency are strongly related, to the point that we can define a partition as a function of latency: the developer can decide the latency threshold beyond which partitioning occurs. By indefinitely retrying communications, one is essentially choosing a CP solution over an AP one, whereas replying right away to a user request means choosing AP over CP, as the data might be stale (inconsistent with the current state). The idea of selecting either CP or AP is a false dichotomy, however, as tuning the time threshold makes it possible to switch from a CP solution to an AP one after a chosen amount of time.

UniquID aims to provide similar flexibility depending on the application domain. This can be easily done by deciding for how long an IoT device will try to download an up-to-date smart contract from the blockchain (i.e., CP behavior) before conceding the resources to the client based on the locally stored smart contract (i.e., AP behavior). In this particular case, the IoT device is a UniquID node that can locally read a smart contract even though it does not store the whole blockchain. The ability to locally read a smart contract means that any network failure happening between the node and the rest of the UniquID network is a partition. Therefore, the CAP theorem applies and a thoughtful decision must be made between consistency and availability.

\section{UniquID Overall Architecture}\label{sec:overall_architecture}
\begin{figure*}[ht]
\center
\includegraphics[trim=0mm 0mm 0mm 0mm,clip,width=0.90\textwidth]{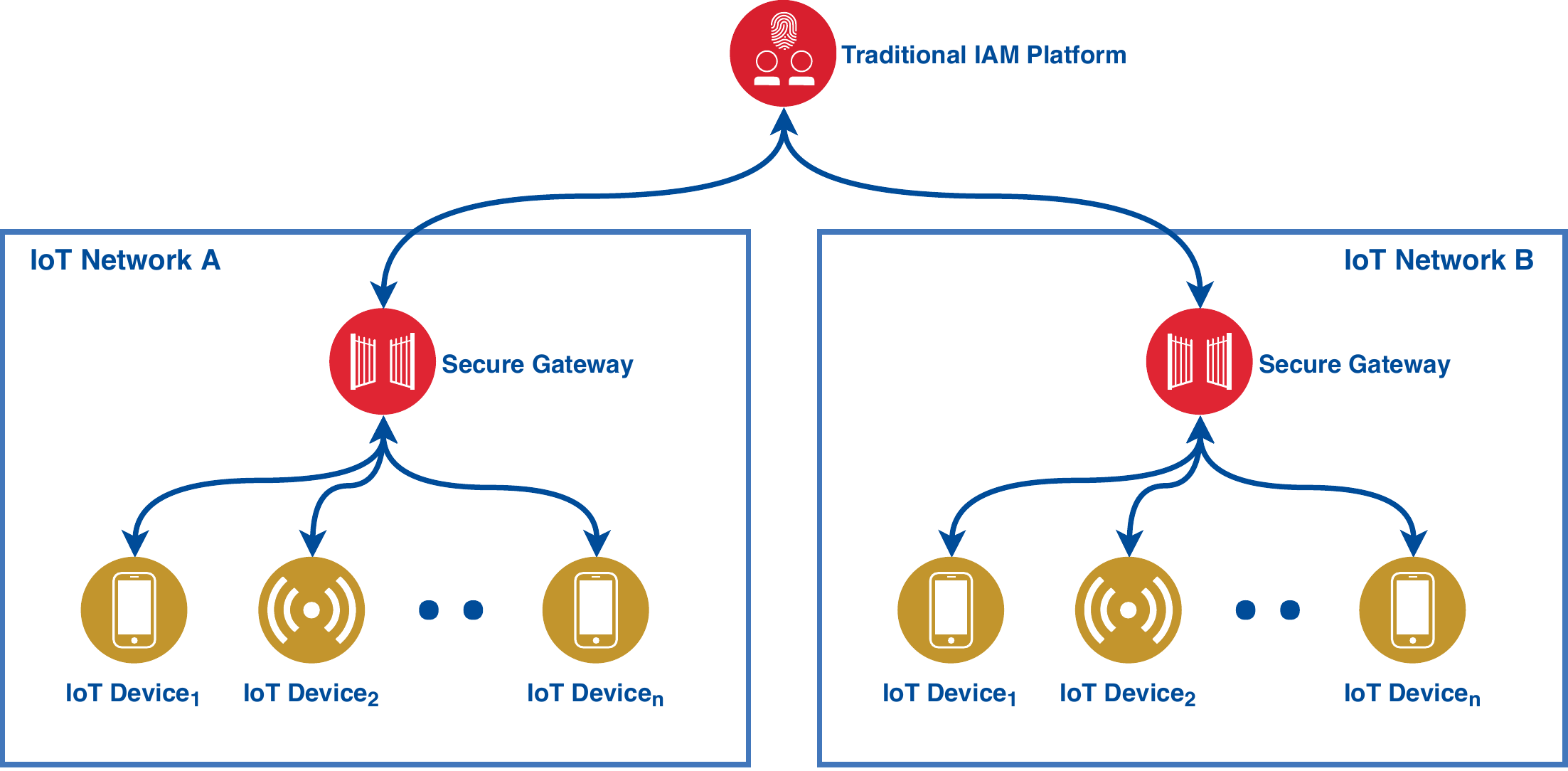}
\caption{Traditional IAM systems impose a hierarchical structure over managed devices. This introduces potential single points of failure in the architecture and strongly impedes direct M2M communications.}
\label{fig:trad_iam}
\end{figure*}

Figure~\ref{fig:trad_iam} shows how different networks interact with each other in a traditional IAM infrastructure. An IoT device in Network A must go through the IAM platform to authenticate on a different IoT network, such as Network B. In the event of network unavailability, a device cannot authenticate, showing that this solution does not provide good availability. 

\begin{figure}[ht]
\center
\includegraphics[trim=0mm 0mm 0mm 0mm,clip,width=0.55\textwidth]{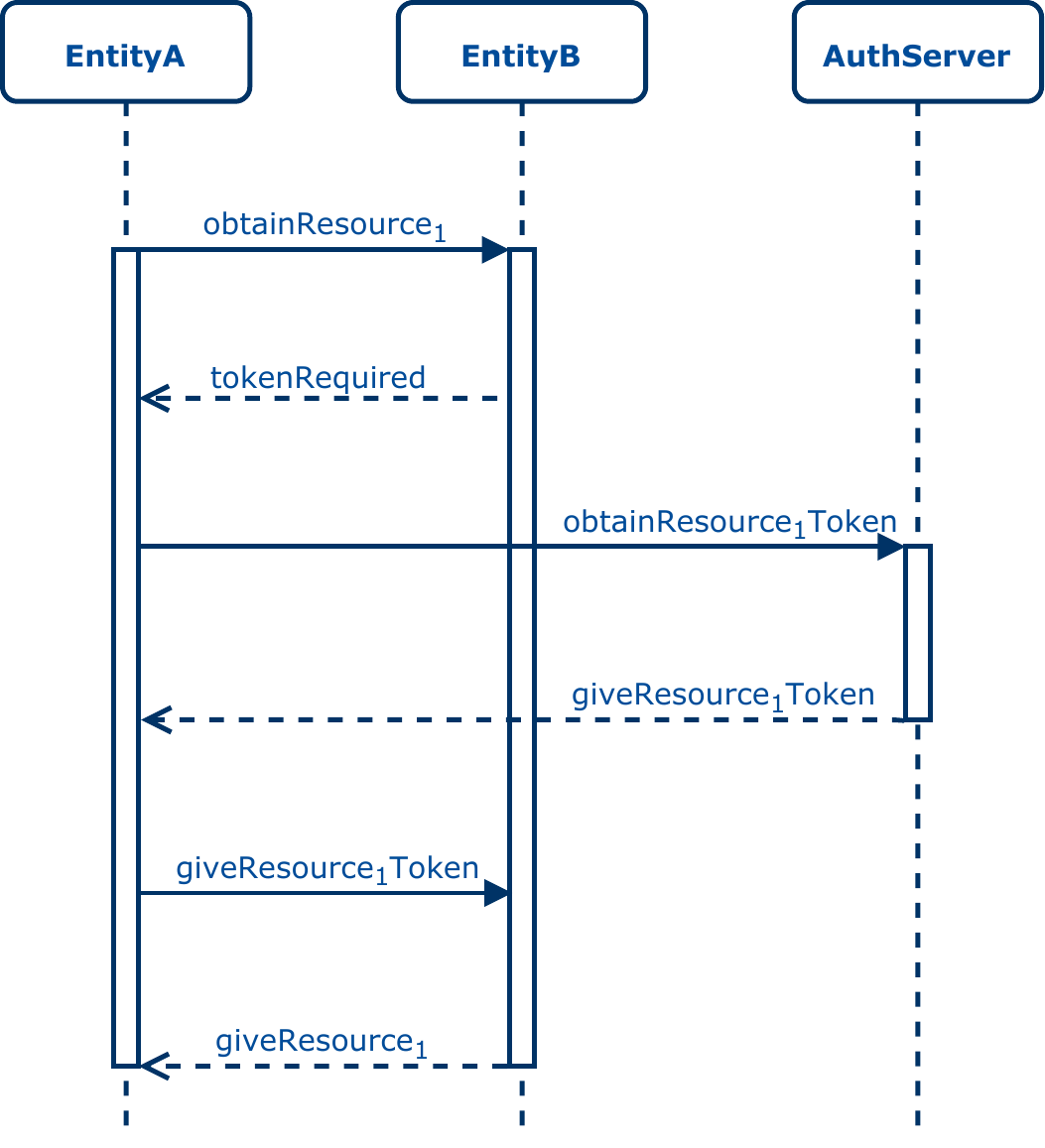}
\caption{With traditional IAM systems, resource negotiation between a requestor (EntityA) and a provider (EntityB) involves an additional exchange between EntityA and an authentication server (AuthServer).}
\label{fig:trad_flow}
\end{figure}

Furthermore, as shown in Figure~\ref{fig:trad_flow}, this kind of architecture involves a considerable number of message exchanges. This overhead decreases the overall responsiveness of the system and means that it cannot ensure availability in the event of network partitioning. Indeed, we can classify a traditional IAM architecture as CP-compliant without any possibility of choosing an AP solution. This system was acceptable in the old days of fixed and stable architectures, but IoT networks require greater flexibility and interoperability.

\begin{figure*}[ht]
\center
\includegraphics[trim=0mm 0mm 0mm 0mm,clip,width=0.90\textwidth]{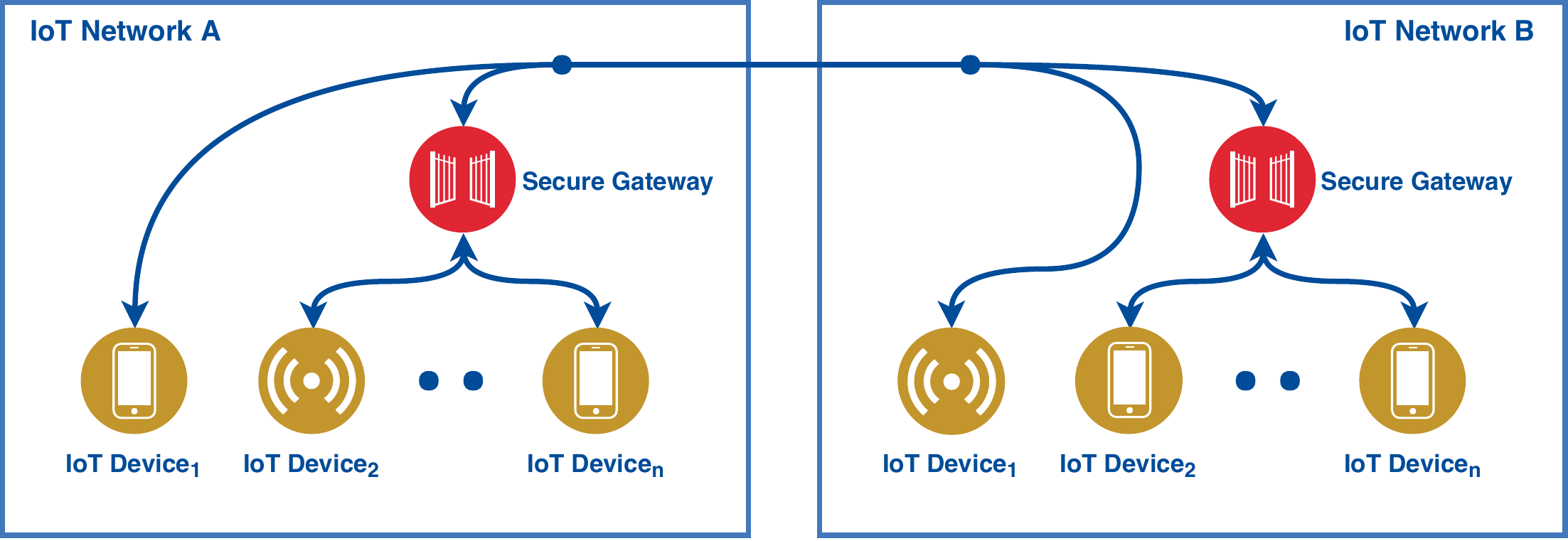}
\caption{UniquID removes the authentication server, which has two important implications. First, the potential single point of failure disappears from the network. Second, scalability greatly improves, as M2M communication is possible and a central authority is no longer essential for negotiation.}
\label{fig:uniquid_iam}
\end{figure*}

A primary goal of UniquID is to replace traditional IAM structures with a less expensive architecture that is more flexible and easier to manage. As shown in Figure~\ref{fig:uniquid_iam}, the system uses an infrastructure where devices can directly authenticate each other without a trusted third party IAM platform, as envisioned by the PGP Web of Trust~\cite{caronni2000walking,WoT00}.

\begin{figure}[ht]
\center
\includegraphics[trim=0mm 0mm 0mm 0mm,clip,width=0.55\textwidth]{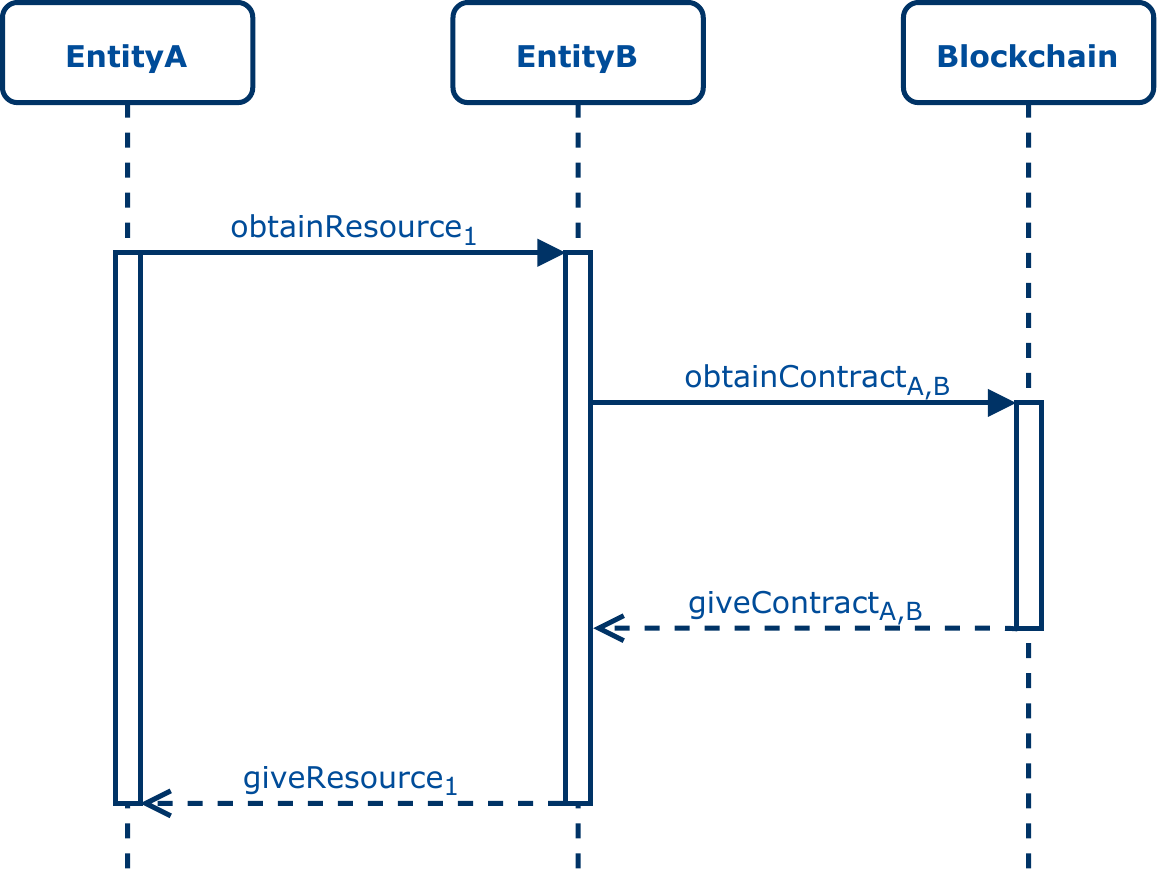}
\caption{UniquID communication is more straightforward than in traditional IAM. EntityA can negotiate for resources directly with EntityB, which can interrogate the blockchain to confirm that the request is valid. But this interrogation is not mandatory, since EntityB can store hash trees and locally confirm certificate validity thanks to the Merkle tree data structure.}
\label{fig:uniquid_flow}
\end{figure}

Figure~\ref{fig:uniquid_flow} shows the UniquID workflow, illustrating two main differences from the IAM setup depicted in Figure~\ref{fig:trad_flow}. First of all, fewer messages are required than in a traditional approach, which entails that UniquID provides better performances than a centralized IAM, assuming that AuthServer and Blockchain have comparable response times. Second, and more important, the entities use smart contracts stored in a tamper-proof public blockchain instead of tokens. This not only ensures higher security but also enables Entity B to store the contract so it can still authenticate Entity A in the event of a network failure. 

This might pose some security issues. Suppose a malicious person M steals Entity A, the laptop of an important CEO. The network administrator immediately revokes all Entity A permissions by issuing a blockchain transaction. If Entity B is instructed to allow authentication without double-checking the currentness of the permission, M might try to gain as much time as possible to cause damage by initiating a network fault that isolates Entity B from the blockchain. This would enable M to authenticate despite the staleness of the Entity A permissions stored within Entity B.

At the same time, the ability to verify a smart contract offline might be invaluable for some other applications. Take as an example a public transportation service. Assume that user U pays for a one-hour ride ticket that allows him to take any city bus line he likes. Assume that the city buses pull this ticket from the blockchain, and temporarily store it locally. After U has done all his errands, he goes to the bus stop and tries to get on the bus, but a network fault occurs. If the buses have been configured to authenticate him without double-checking the current status of the ticket on the blockchain, U will be able to board the bus. If, on the other hand, the buses need to perform a double-check, U will be left waiting until the network fault is resolved, which might take a long time. Again, a malicious user M might isolate the bus he is boarding and extend his ticket for some time. However, this is a small economic risk that the company would be willing to take given the more serious consequences that a network fault would have on all of its  transportation infrastructure.

Each scenario has different requirements. One of the strengths of UniquID is that the end users can decide whether a CP system or an AP system is best suited to their needs.

\subsection{Imprinting Ceremony}
With the goal of excluding any kind of PKI, we envision an approach similar to the Web of Trust paradigm~\cite{caronni2000walking,WoT00} for the node initialization phase. In UniquID, the imprinting ceremony resembles a PGP key-signing party. To be enrolled in UniquID, every node must be initialized, which happens in different phases. To minimize the chances of man-in-the-middle (MitM) attacks, the following steps should take place as soon as possible in the production chain.

First of all, each node generates its own public key, which is directly (i.e., not through the Internet) exchanged with other devices in the same local network. After this phase is complete, the identities are stored in a blockchain through the imprinting process. An \textit{imprinter} is a node designated to collect all the new identities, and forward them to the blockchain of choice.

In details, the imprinter is appointed to perform some critical tasks. First, the imprinter generates a special smart contract, the imprinting contract (IC), which links the device to the public key of its administrator. Once the generation is done, the imprinter collects the ICs and announces them to the blockchain, in order to ensure data immutability. From this moment on, these enrolled devices are able to interact with the UniquID infrastructure. If the device is transferred to another entity (e.g., sold to a customer), the administrator (e.g., the manufacturer) signs a contract and transfers the administrative rights by replacing its public key with that of the new owner. Ideally, due to the critical role of the imprinter, the manufacturer deploys a number of different imprinting nodes, to both speed up the imprinting process, and to avoid single point of failures in the architecture.

\begin{figure*}[ht]
\center
\includegraphics[trim=0mm 0mm 0mm 0mm,clip,width=0.95\textwidth]{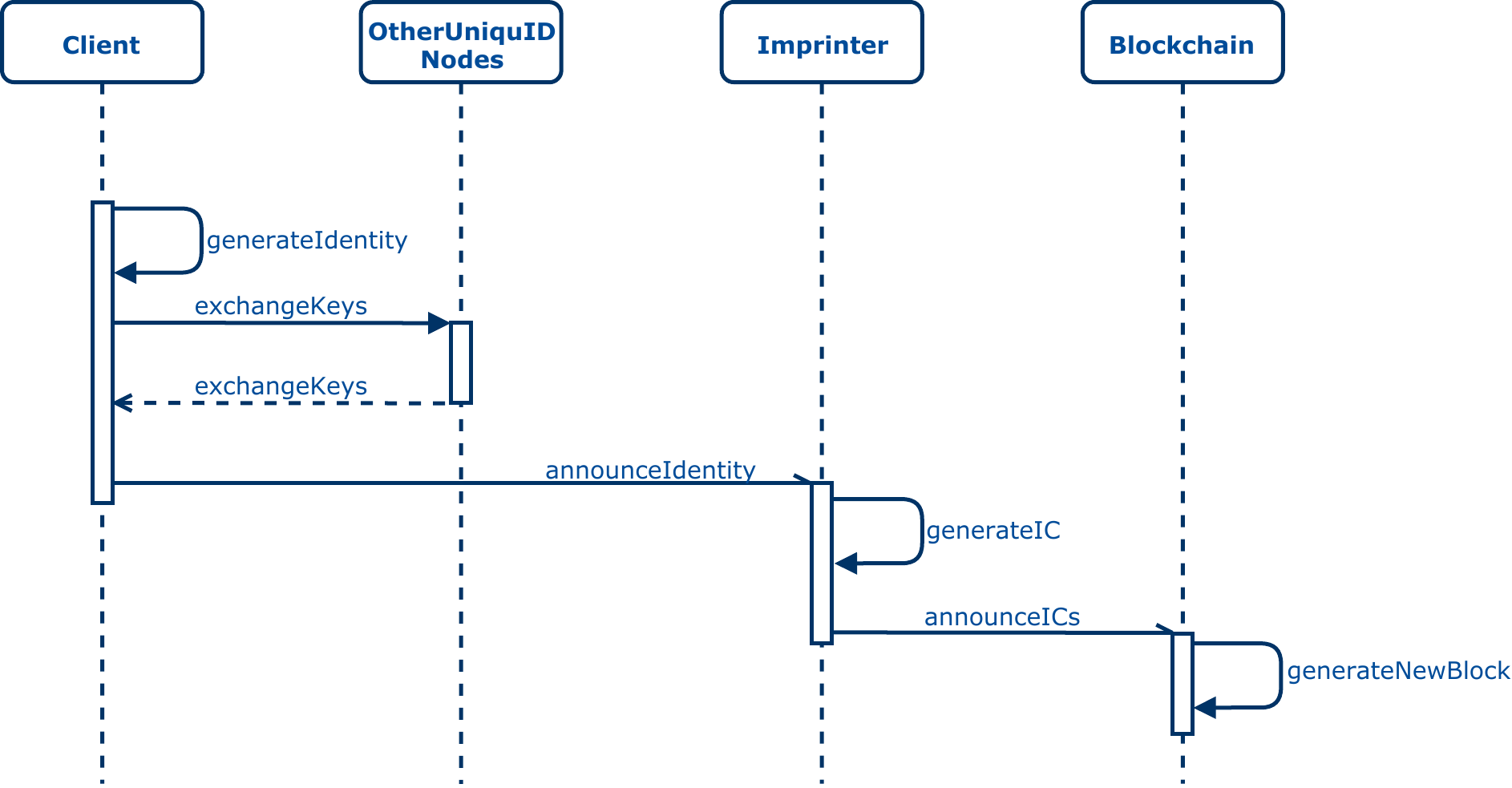}
\caption{Sequence diagram that depicts a standard imprinting ceremony.}
\label{fig:imprinting_sequence}
\end{figure*}

In Figure~\ref{fig:imprinting_sequence}, we show the main tasks performed during the imprinting phase. After that a node has created its identity (i.e., the public key), it announces it to the imprinter which is appointed to create the IC and to communicate it to the blockchain. At that point, the IC will go under the underlying blockchain processes, until it will be included in a new block and become almost immutable.

\section{Experimental Evaluation}\label{sec:evaluation}
In this section, we present the experimental results of a UniquID enrolment instance, executed on a cloud service. Even though we measured the performance of the identity generation phase, as well as the imprinting phase, our goal is not to assess the performance of such instance. Indeed, the system scalability is tightly tied to the utilized resources, and the small setup we used for our experiments cannot represent a fully operating UniquID network. The aim of these experiments is to show that our solution is feasible, and that it successfully stores immutable identities on a public blockchain. On a larger scale, this would be enough to make PKIs, passwords, and certificates unnecessary.

To perform our experiments, we created 7 parallel clients, designated to generate 1000 virtual IoT identities and communicate such identities to 1 imprinting node. Every identity is announced as soon as it is created, and the receiving imprinting node is appointed both to create the IC transactions and forward them to the Litecoin Testnet blockchain. Imprinter and clients ran over AWS T2.Micro instances, burstable performance instances equipped with 1 Gb of RAM. As a communication protocol between the imprinter and the clients, we used MQTT (Message Queue Telemetry Transport), an ISO standard (ISO/IEC PRF 20922) publish-subscribe-based messaging protocol~\cite{mqtt}, designed for lightweight communications.

As aforementioned, we chose Litecoin as a storing public blockchain, which ensures one mined block every $2.5$ minutes. Considering that a Litecoin block is 1 Mb and that a UniquID transaction is 400 byte, this design choice entails a theoretical upper bound of 2500 enrolled devices per $2.5$ minutes, or 1000 devices per minute. More in general, we can define the theoretical upper bound of enrolments per minute with the following equation:

\begin{equation}
\frac{Block~Size~(bytes)}{400~(bytes)~\cdot~Average~Mining~Time~(m)}\,,
\end{equation}
where 400 bytes are the size of a UniquID transaction, and the other parameters depend on the underlying blockchain.

\subsection{Identity Generation}
In the first evaluation phase, we measured the time required to generate an identity on a virtual client. As shown in Figure~\ref{fig:generating_id}, on average it took $6.61 \pm 0.03$ milliseconds to generate an identity.


As stated before, in our experiments we delegated the generation task to the virtual clients. In an ideal scenario, the IoT devices would create and announce their identity themselves, therefore we performed an additional experiment, in order to evaluate the generation time on a low power device. The device we used was equipped with an \textit{arm926ejste} CPU, $256$ MB RAM, and a \textit{mlinux 3.3.6} OS. On this IoT device, the generation process took $627$ milliseconds, 44.7\% of CPU and 0.4\% of RAM. The footprint left on the device storage was $6.70$ MB.

While the generation time of $627$ milliseconds is substantially higher than the one obtained with a virtual client ($6.61 \pm 0.03$ milliseconds, as showed in Figure~\ref{fig:generating_id}), generating an identity well under $1$ second is still an excellent result. Moreover, the resources used by the whole process are reasonably low and the resulting footprint is small enough to fit into any modern IoT device.

\subsection{Imprinting}
\begin{figure*}[ht]
    \centering
    \begin{subfigure}[t]{0.48\textwidth}
        \centering
        \includegraphics[trim=0mm 0mm 0mm 0mm,clip,width=1\textwidth]{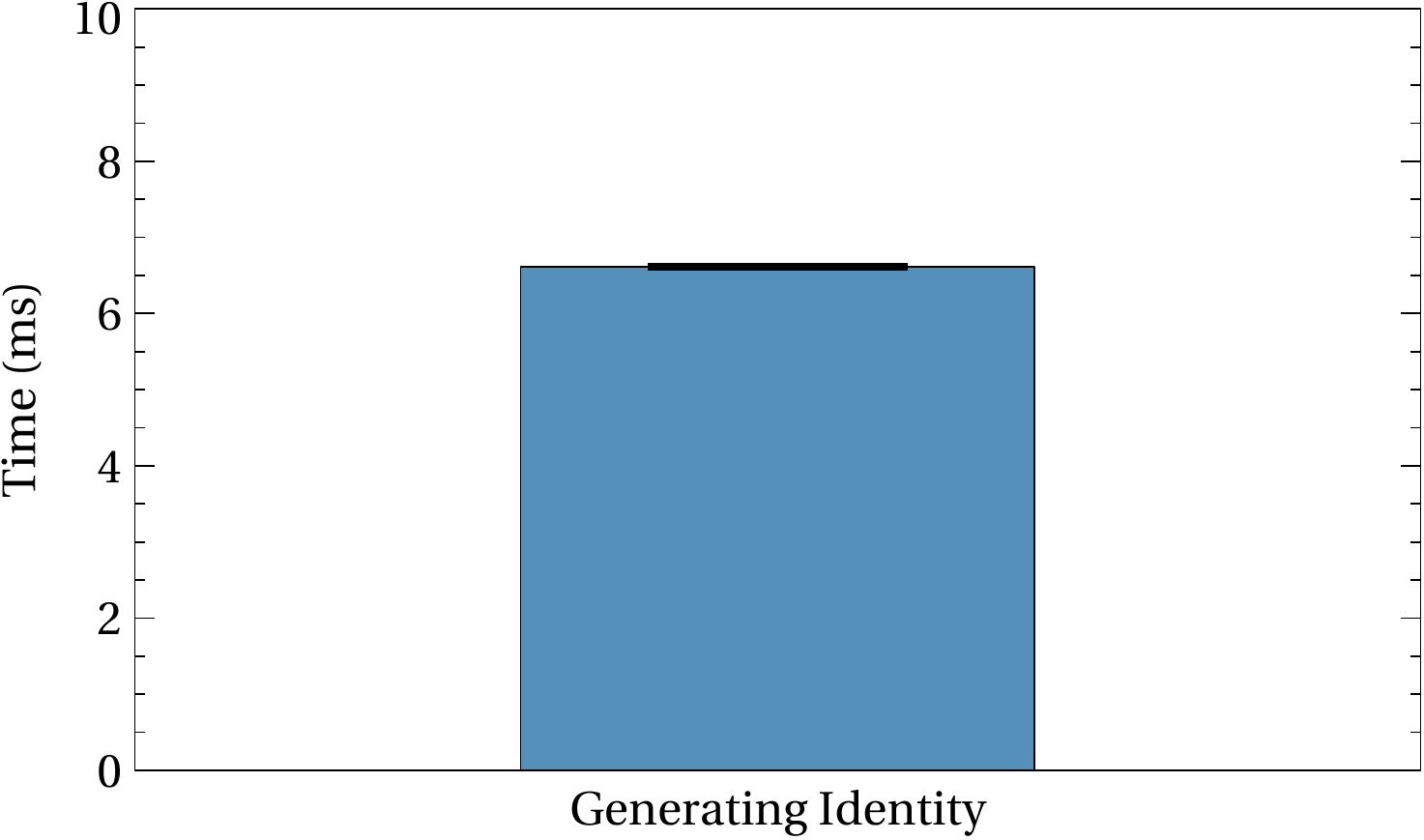}
        \caption{Time is in \emph{milliseconds}, error bar shows the Standard Error (SE).}
        \label{fig:generating_id}
    \end{subfigure}%
    ~ 
    \begin{subfigure}[t]{0.48\textwidth}
        \centering
        \includegraphics[trim=0mm 0mm 0mm 0mm,clip,width=1\textwidth]{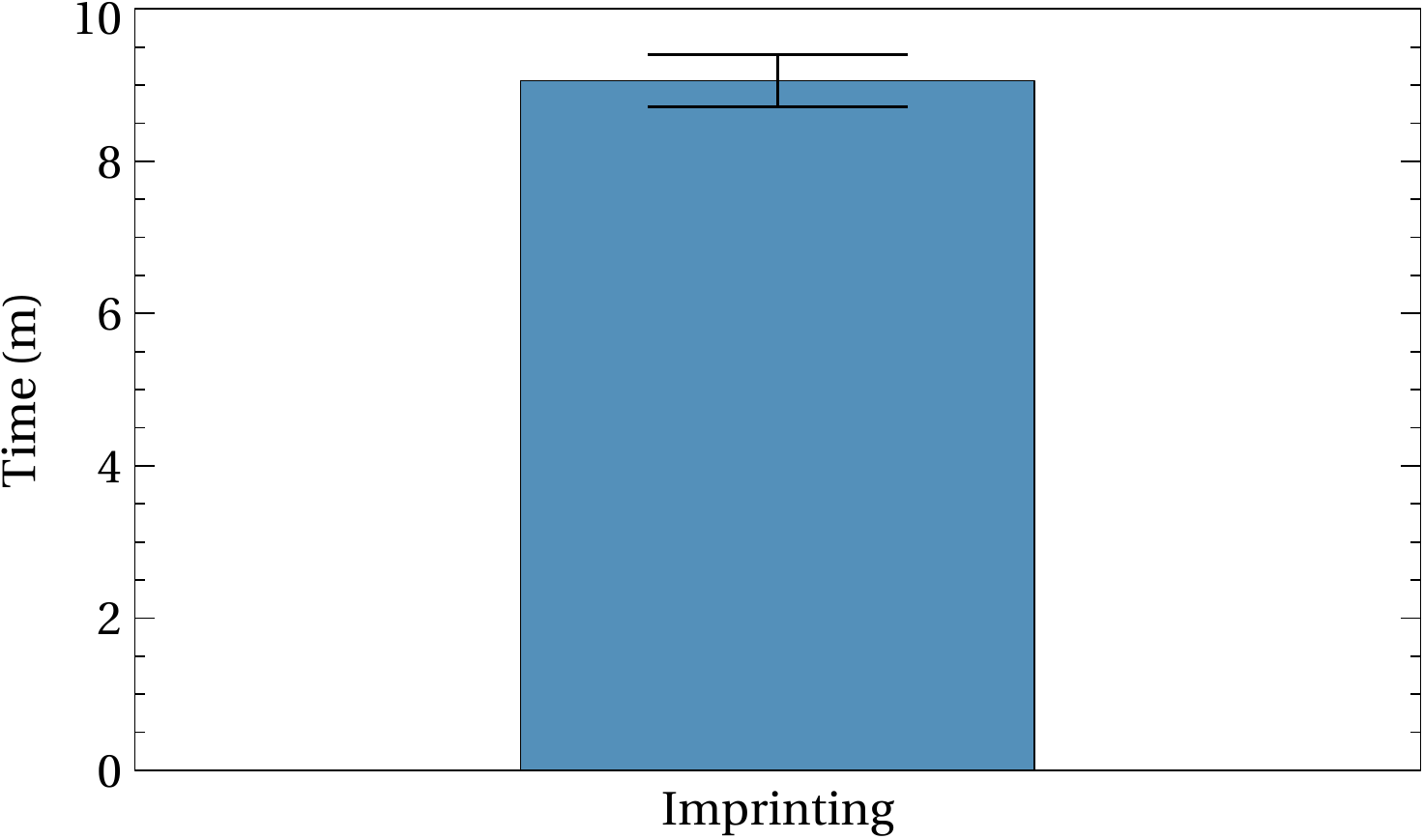}
        \caption{Time is in \emph{minutes}, error bar shows the Standard Error (SE).}
        \label{fig:imprinting}
    \end{subfigure}
    \caption{Average times for generating and imprinting identities, respectively.}
\end{figure*}

As a second part of our experiment, we analysed and measured the necessary time to imprint new identities on a public blockchain. As aforementioned, our experimental setup is composed by 6 clients that created 1000 identities, and announced them to a single imprinter node. Again, the imprinter has the role to create an IC transaction for each identity and submit all the transactions to the blockchain. In Figure~\ref{fig:imprinting}, we show that the average time to imprint one identity is $9.06 \pm 0.35$ minutes.

Our experiments clearly show that the imprinting took a considerably longer time than the generation of the identities. The former took minutes, whereas the latter took milliseconds.

\subsection{Summing Up: Enrolment}

\begin{figure}[!ht]
\center
\includegraphics[trim=0mm 0mm 0mm 0mm,clip,width=0.5\textwidth]{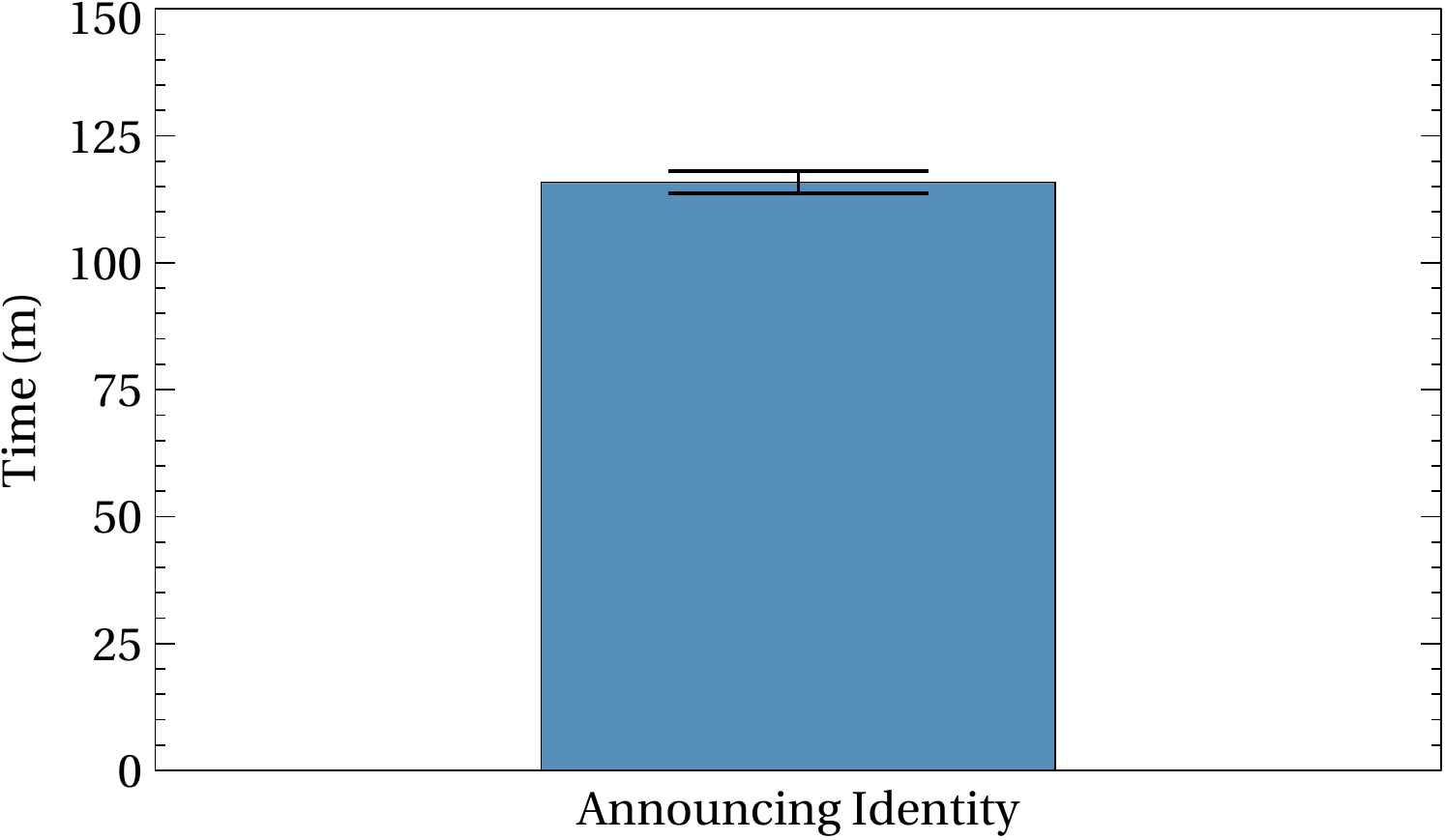}
\caption{Average time to announce one identity. Announce time is in minutes, error bar shows the Standard Error (SE).}
\label{fig:announcing}
\end{figure}

Our experiments show that, in total, it took around $4.5$ hours to automatically enrol (i.e., create, announce, and imprint) 1000 identities, without any human intervention. The reason why this practical result is considerably under the theoretical upper bound of 1000 enrolments per minute, is that our virtual clients on Amazon AWS could open no more than 5 concurrent sockets, per each. This resulted in a bottleneck, where already generated identities were announced with considerable delays. Figure~\ref{fig:announcing} shows the magnitude of this bottleneck: on average, it took $115.86 \pm 2.19$ minutes to announce an identity to the imprinter, which is a significant amount of time compared to the time for identity generation ($6.61 \pm 0.03$ milliseconds) and imprinting ($9.06 \pm 0.35$ minutes).

\noindent\textbf{Internal Experiment.} In order to further investigate this communication bottleneck, we ran some internal tests with exactly the same network setup. These tests showed that we can easily run up to 50 concurrent threads per MQTT client, overcoming the communication bottleneck and increasing our capacity to 400 enrolments per minute. Still, at the present stage UniquID does not enable us to get near the theoretical upper bound, but this is purely due to software limitations that are currently addressed.

To sum up, our experiments successfully show that is possible to automatically enrol IoT devices in a blockchain-based IAM. In practice, this means that enrolled devices no longer need to rely upon PKIs, centralized CA authorities, certificates, nor passwords. Moreover, our experiments lead to immutable identities, publicly verifiable by anyone on the Litecoin Testnet~\footnote{For instance, the first enrolment is verifiable at: \url{https://testnet.litecore.io/tx/feac5a1dc645c701ea17ceb1657541d7094fbb43f18749cd9bd8 a54014bd0197}}. Once showed that our approach works in practice, the next step will be to optimise the performance to get as close as possible to the theoretical upper bound.

%
%
%

\section{Case Studies}\label{sec:case_studies}
In this section, we present a couple of case studies to highlight UniquID strengths and variety of applications.

\subsection{Smart Vehicle}
Let us assume a connected vehicle with 3G cellular capability and a mobile app designed to unlock the doors or start the engine. In a typical scenario, the smartphone app connects to a RESTful cloud service responsible for authenticating the user and forwarding the command to the vehicle, leveraging an encrypted session pushed through an available 3G connection. This architecture presents three main issues:
\begin{itemize}
\item In the absence of 3G signal coverage, the cloud service cannot perform any remote command and control of the vehicle, leaving the user locked out~\cite{boyle_2017}.
\item In large-scale deployments, the cloud service becomes the main bottleneck of the system, introducing latency and potential downtime during peak hours.
\item Every vehicle-side application is exposed to Internet connectivity and thus must be maintained against zero-day vulnerabilities and malware and ransomware attacks~\cite{greenberg_2015}.
\end{itemize}

\begin{figure}[ht]
\center
\includegraphics[trim=0mm 0mm 0mm 0mm,clip,width=0.6\textwidth]{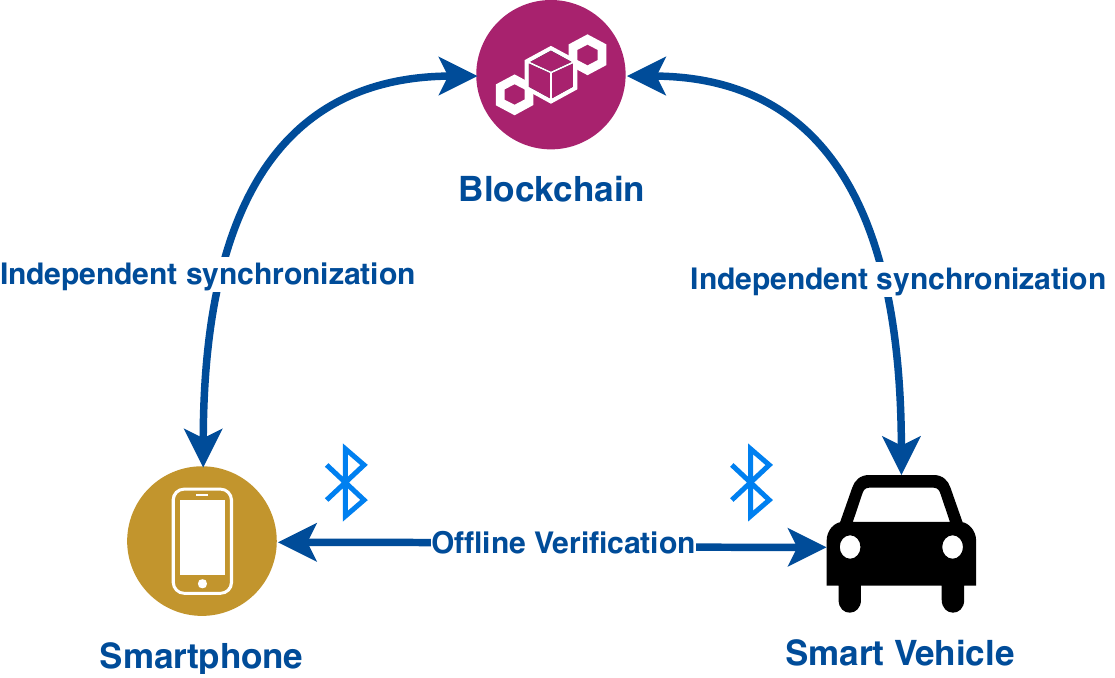}
\caption{UniquID enables M2M offline verification of smart contracts, which makes possible to start a loaned smart vehicle even in case of network failure.}
\label{fig:smart_vehicle}
\end{figure}

UniquID allows the two endpoints (i.e., the smartphone and vehicle) to independently synchronize with the ledger, leveraging the hashcash PoW cost-function~\cite{back2002hashcash} to verify the integrity of a new block. Moreover, transactions stored as Merkle tree leaves allow efficient verification of the veracity of a new authorization~\cite{nakamoto2008bitcoin}.
Once this information is locally stored on both ends, handshake and authorization are performed through low-energy and proximity protocols (e.g., BLE) without relying on Internet connectivity or remote command-and-control services, as previously shown in Figure~\ref{fig:uniquid_flow}. Figure~\ref{fig:smart_vehicle} shows the resulting architecture.


\subsection{Industrial Sensor Network}
Consider an industrial Internet application where a large network of remotely installed, battery-operated sensors collect time series data in a low-connectivity environment. These sensors do not have cellular signal coverage, and satellite uplink is not a feasible option. The data from these sensors is periodically "harvested" by human operators through a portable device, such as a laptop or rugged tablet, using an ad-hoc Wi-Fi connection to the sensors. This architecture presents two main challenges:
\begin{itemize}
\item IAM for the sensors is achieved through Wi-Fi passwords, which require complex maintenance and periodic rotation (e.g., in case of operator change).
\item Multi-tenancy scenarios require an additional layer of authentication at the application level, which needs to manage passwords stored on the sensor itself.
\end{itemize}

UniquID approach pushes the storage of these credentials onto the distributed ledger, enabling a remotely controlled, asynchronous IAM solution. A typical implementation leverages a central orchestration interface that sends microtransactions containing the ACL between one sensor and its authorized users on the distributed ledger. As a result, no passwords are needed: combining the ledger pseudonymous identity from the wallets (installed on the operator's device and the sensor), and blockchain-stored transactions that contain the ACLs, a sensor can recognize the operator's device and provide the data that the user is authorized to collect. 

\begin{figure}[ht]
\center
\includegraphics[trim=0mm 0mm 0mm 0mm,clip,width=0.8\textwidth]{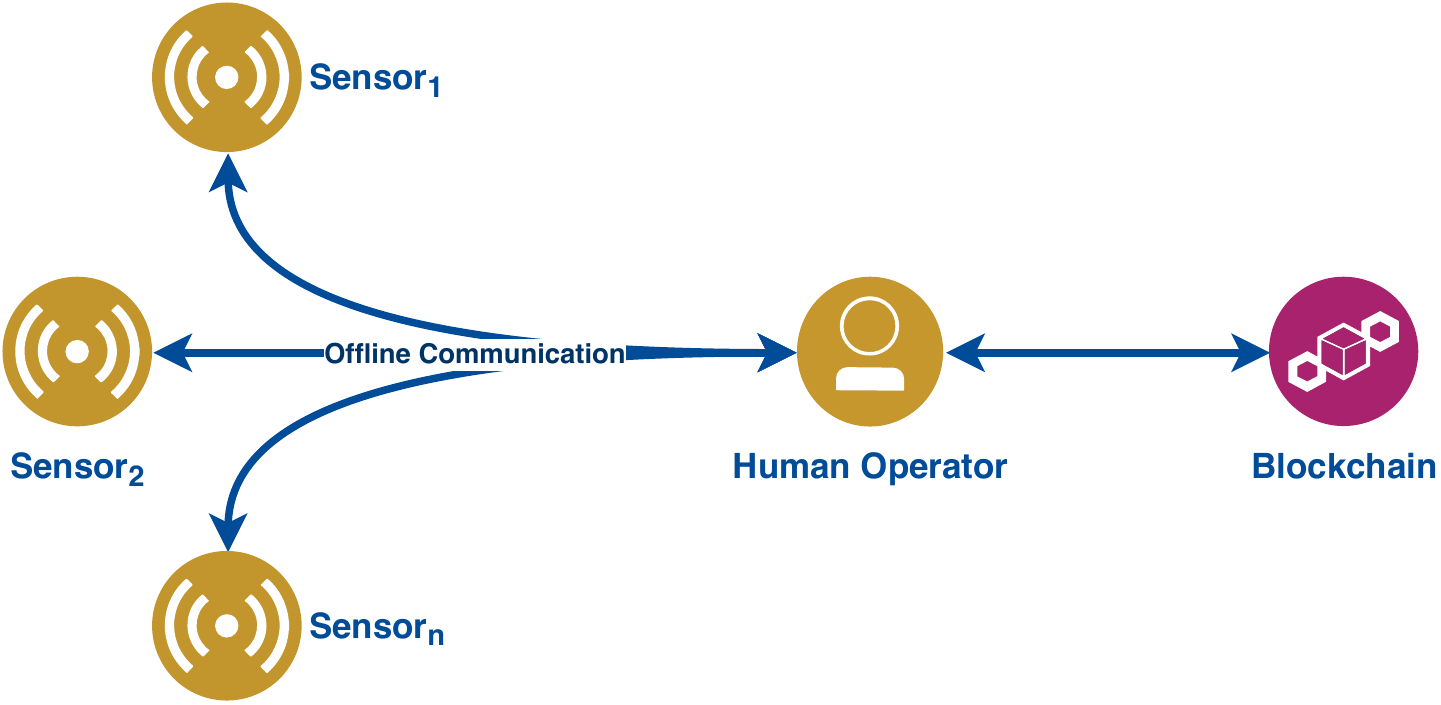}
\caption{UniquID enables a human operator to authenticate against non-connected sensors. Furthermore, the user dispatches to the sensors the latest blocks of the shared ledger.}
\label{fig:industrial_sensor_network}
\end{figure}

Furthermore, the operator acts as the dispatcher of the most recent blocks on the ledger. PoW is sufficient to verify the integrity of any received block, reducing the risk of forged blocks (which would require network consensus) and removing the need for a trusted session between peers. In this way, even if always disconnected from the Internet, sensors can be kept up to date with the latest authorizations stored on the ledger. Figure~\ref{fig:industrial_sensor_network} shows the resulting architecture.

\section{Conclusion}\label{sec:concl}
In this paper, we first briefly covered why current IAM systems and PKIs are too cumbersome for the challenges posed by the IoT paradigm. We also showed how the CAP theorem strongly applies to blockchain-based IAMs, and how developers can independently decide how their applications react to network partitions.

We illustrated the overall UniquID architecture and how its SDK aims to simplify both the trust architecture as a whole and resource negotiation flows. We also discussed in detail how the backbone architecture of a blockchain-based IAM can be designed, as well as the benefits and drawbacks of each solution.

We provided an overview of the essential features that devices need to implement to participate in UniquID, as well as how the initial imprinting ceremony of such devices happens. We performed a proof-of-concept experiment which led to generating and imprinting 1000 identities on the Litecoin Testnet, virtually immutable and publicly verifiable by anyone. Last, but not least, we provided two relevant case studies that showed how UniquID could help unlock the potential of IoT. In future works we will conduct extensive experiments to assess the robustness and responsiveness of UniquID under a number of adverse conditions, such as man-in-the-middle and denial-of-service attacks.

Having shown the feasibility of our approach through practical experiments, future work will focus on optimising the performance of the system to get as close as possible to the theoretical upper bound of 1000 enrolled devices per minute.

\subsection*{Acknowledgements}
We would like to thank Charles Kozierok for his help to proofread the manuscript.

%
%
%
\bibliographystyle{splncs04}
\bibliography{bibliography}
%
%
%
%
%
\end{document}